\documentclass[10pt, aps, prl, amssymb, amsmath, amsfonts, showpacs, floatfix, twocolumn, twoside, a4paper, superscriptaddress, longbibliography, nofootinbib]{revtex4-2}

\usepackage{mathrsfs}
\usepackage{siunitx}
\usepackage{comment}
\usepackage{soul}
\usepackage[T3,T1]{fontenc}
\DeclareMathAlphabet{\mathpzc}{OT1}{pzc}{m}{it} 
\usepackage{mathrsfs} 
\usepackage{bm} 
\usepackage{microtype} 
\def\MT@register@subst@font{
	\MT@exp@one@n\MT@in@clist\font@name\MT@font@list
	\ifMT@inlist@\else\xdef\MT@font@list{\MT@font@list\font@name,}\fi}
\makeatother
\usepackage{orcidlink}
\usepackage{multirow}
\usepackage{booktabs} 
\usepackage{array} 
\usepackage{tabularx}
\usepackage{braket} 

\newcolumntype{C}{>{\centering\arraybackslash}X}

\usepackage{xcolor}
\usepackage{hyperref} 
\usepackage{nicefrac} 

\hypersetup{
	colorlinks=true,    
	citecolor=blue,     
	linkcolor=red,      
	urlcolor=blue,     
}

\usepackage{graphicx} %
\graphicspath{{./figures/}}
\usepackage{xspace}

\newcommand{\dd}{\mathrm{d}}
\newcommand{\env}{\mathrm{e}}
\newcommand{\GRAPPA}{Gravitation Astroparticle Physics Amsterdam (GRAPPA),\\ University of Amsterdam, 1098 XH Amsterdam, The Netherlands}

\DeclareSymbolFont{mathtx}{OML}{txmi}{m}{it}
\DeclareMathAlphabet\mathbfcal{OMS}{cmsy}{b}{n}
\DeclareMathSymbol{v}{\mathalpha}{mathtx}{118}

\begin{document}
\title{Fully Relativistic Treatment of Extreme Mass-Ratio Inspirals\\in Collisionless Environments}
\author{Rodrigo Vicente}
\email{r.l.lourencovicente@uva.nl}	
\affiliation{\GRAPPA}
\author{Theophanes K. Karydas}
\affiliation{\GRAPPA}
\author{Gianfranco Bertone}
\affiliation{\GRAPPA}

\date{\today}

\begin{abstract}
Future mHz gravitational wave (GW) interferometers will precisely probe massive black hole environments, such as accretion disks, cold dark matter overdensities, and clouds of ultralight bosons, as long as we can accurately model the dephasing they induce on the waveform of extreme mass-ratio inspirals (EMRIs). Most existing models rely on extrapolations from Newtonian results to model the interaction of the small black hole in an EMRI system with the environment surrounding the massive black hole. Here, we present a fully relativistic formalism to model such interaction with collisionless environments, focusing on the case of cold dark matter overdensities, like ``spikes'' and ``mounds''. We implement our new formalism in the \texttt{FastEMRIWaveforms} framework, and show that the resulting waveforms are significantly different from those based on a Newtonian treatment of environmental effects. 
Our results indicate that a fully relativistic treatment is essential to capture the environmental dephasing of GW signals from EMRIs accurately.
\end{abstract}

\maketitle

{\em Introduction.---}
%
Next-generation GW detectors like LISA~\cite{LISA:2022kgy}, TianQin~\cite{TianQin:2015yph}, and DECIGO~\cite{Kawamura:2020pcg} will allow us to probe the environments where compact binaries form, evolve, and merge~\cite{Cardoso:2019rou, CanevaSantoro:2023aol, Bertone:2024rxe, Zwick:2025wkt, LISA:2024hlh}.
In particular, these mHz detectors will allow us to track individual extreme mass-ratio inspirals (EMRIs) over hundreds of thousands of orbital cycles, spanning months to years, opening up new opportunities for astrophysics~\cite{Barausse:2007dy, Kocsis:2011dr, Yunes:2011ws, Barausse:2014tra, Tamanini:2019usx, Vicente:2019ilr, Toubiana:2020drf, Derdzinski:2020wlw, Zwick:2021dlg, Sberna:2022qbn, Zwick:2022dih, Vijaykumar:2023tjg, Duque:2024mfw, Roy:2024rhe, Spieksma:2025wex, Copparoni:2025jhq, Duque:2025yfm} and fundamental physics~\cite{Barack:2018yly, Bertone:2019irm, Annulli:2020lyc, Kavanagh:2020cfn, Baumann:2021fkf, Maselli:2021men, Coogan:2021uqv, Vicente:2022ivh, Traykova:2023qyv, Duque:2023seg, Cardoso:2021wlq, Cardoso:2022whc, Spieksma:2024voy, Speeney:2024mas, Pezzella:2024tkf, Gliorio:2025cbh, Cole:2022yzw, 2307.16093, Duque:2023seg, 2501.09806, Tomaselli:2023ysb, Mitra:2023sny, Rahman:2023sof, Karydas:2024fcn, Kavanagh:2024lgq, Spieksma:2024voy, Blas:2024duy, Miller:2025yyx, Boskovic:2018rub, Boskovic:2024fga}.
The scientific potential of EMRIs hinges critically on the accuracy of their theoretical modeling. While these systems are inherently relativistic~\cite{2410.17310}, their interactions with surrounding matter are often approximated using Newtonian extrapolations, sometimes supplemented by special relativity corrections (e.g.,~\cite{Barausse:2007ph, Speeney:2022ryg, Mitra:2025tag}); a few notable exceptions are, e.g.,~\cite{1010.0758, 1010.0759, Duque:2025yfm} for accretion disks,~\cite{2307.16093, Duque:2023seg, 2501.09806} for ultralight bosons, and~\cite{Correia:2022gcs} for non-gravitational forces. The inaccurate modeling of environmental effects might lead to false violations of general relativity~\cite{Garg:2024qxq}, systematic biases~\cite{Roy:2024rhe}, or, in extreme cases, to missing signals~\cite{Roy:2024rhe}.  

In this Letter, we develop a relativistic framework to model environmental effects on EMRIs in collisionless media. We then focus on dark matter (DM) spikes as a key example, assessing the significance of relativistic effects in such systems. Our results demonstrate that a fully relativistic treatment is necessary to accurately model these interactions. We adopt geometric units ($c = G = 1$) and use the mostly positive metric signature. The symbol $\overset{*}{=}$ denotes evaluation in a particular frame. \\

{\em Adiabatic evolution.---}
%
The adiabatic modeling of EMRIs~\cite{gr-qc/0504015, 2410.17310} relies on a two-timescale expansion~\cite{0805.3337, 2006.11263}, which separates the short orbital timescale, $T_{\rm orb} \sim M$, over which dissipative effects are evaluated, from the long radiation-reaction timescale, $T_{\rm rad} \sim M/q$, over which the orbit evolves; here, $M$ is the mass of the primary black hole (BH), and $q \equiv m/M \ll 1$ is the system’s mass ratio. In this framework, the system is described by a sequence of geodesics evolving at a rate determined by the orbit-averaged dissipative fluxes of GW radiation. The adiabatic approximation is equivalent to the leading-order dissipative effects of the first-order gravitational self-force~\cite{1805.10385, Pound:2021qin_selfforce}, neglecting conservative effects and (averaged-out) higher-order dissipative effects~\cite{gr-qc/0509122}.

We build on this to develop a relativistic formalism to account for the effects of collisionless environments on EMRIs at adiabatic order (similarly to the approach taken in~\cite{2410.17310} for clouds of ultralight scalars). We consider a background spacetime described by the Kerr metric, neglecting the gravitational backreaction of the environment, which is tiny around EMRIs for most astrophysical systems, as both the enclosed mass within the orbit and the environment's compactness are minuscule (more on that later). A geodesic $\xi$, with four-velocity $\bm{u}\equiv \dd \xi/\dd \tau$ and proper time~$\tau$, is uniquely determined by its energy, angular momentum around the axis of symmetry (i.e. the primary's spin axis), and Carter's constant defined by\footnote{To completely fix the geodesic a point~$\xi(\tau_0)$ must be supplemented.}
\begin{equation}
	\varepsilon\equiv -\bm{u}\cdot \bm{\partial}_t\,, \quad l_z\equiv \bm{u}\cdot \bm{\partial}_\varphi\,, \quad C\equiv K(\bm{u},\bm{u})\,,
\end{equation}
here normalized by the companion's rest mass, where $\bm{\partial}_t$ and $\bm{\partial}_\varphi$ are, respectively, the timelike and the azimuthal Killing vectors of Kerr, and $K(\cdot,\cdot)$ is a second order Killing tensor of Kerr defined as in~\cite{Wald:1984rg}. 

The flux of GW radiation to infinity and down the primary's event horizon can be computed by solving the Teukolsky equation, as detailed in~\cite{Cutler:1994pb, gr-qc/9910091, gr-qc/0509101}. Their contribution to the secular evolution of the energy and angular momentum along the primary's spin can be found directly from global conservation laws and, while more indirectly, the evolution of the Carter's constant can also be determined by similar techniques~\cite{gr-qc/0302075, gr-qc/0504015, gr-qc/0505075, gr-qc/0506092}. 
A \linebreak four-acceleration $\bm{a}$ defined over $\xi$ describing the interaction of the companion with a local environment results in the additional contribution
\begin{equation}\label{eq:rates0}
	\dot{\epsilon}_\env = -{u^t} \bm{a} \cdot \bm{\partial}_t \,, \;\,\,\, \dot{l}_{z,\env} = {u^t} \bm{a} \cdot \bm{\partial}_\varphi\,, \;\,\,\, \dot{C}_\env\equiv 2 u^t K(\bm{a},\bm{u})\,,
\end{equation}
where the dot denotes the derivative with respect to the Boyer-Lindquist time $t$. These, together with the contribution of the GW fluxes, can then be orbit-averaged to compute the adiabatic evolution of the EMRI. The key question is then: how to compute the four-acceleration from a (local) interaction with an environment within general relativity?\\

{\em Collisionless environment.---}
%
We consider a kinetically supported collisionless medium, under the gravitational influence of a supermassive BH---this could be, for instance, a cold DM overdensity~\cite{Gondolo:1999ef, Sadeghian:2013laa, Ferrer:2017xwm,Bertone:2024wbn} or a radiatively inefficient accretion disk~\cite{Cremaschini:2011, Hoshino:2015cka, Vos:2024loa}. 
Such media is described by a distribution function $F(x,\bm{p})$, representing the invariant number $\dd N$ of particles contained in a spacelike volume element $\dd V$ at $x$, and whose four-momenta $\bm{p}$ lie within the corresponding mass-shell three-surface element $\dd P$ in momentum space, which is constrained by $\bm{p}\cdot \bm{p}=-\mu^2$, with $\mu$ the rest mass of the particles. This distribution function is independent of $\dd V$, when defined through $\dd N\equiv F(x,\bm{p})(-\bm{p}\cdot \bm{n})\dd V \dd P$, where $\bm{n}$ is the future timelike unit normal to $\dd V$~\cite{Lindquist:1966igj}.
In the following, it will be convenient to introduce (local) normal coordinates at $x$ associated to the orthonormal frame $(\bm{u},\bm{e}_a)$. In the induced coordinates in momentum space, the volume element is $\dd P = (1/p^0) \, \dd p^1 \dd p^2 \dd p^3$.

From Jeans' theorem~\cite{Jeans:1915}, the distribution function of a steady-state, collisionless system depends on the phase-space coordinates only through the integrals of motion. Assuming a relaxed steady-state environment, we take $F(x,\bm{p})= \mu^{-3} f(\mathcal{E},\mathcal{C},\mathcal{L}_z)$, where $\mathcal{E}(x,\bm{p})$, $\mathcal{C}(x,\bm{p})$, and $\mathcal{L}_z(x,\bm{p})$ are, respectively, the energy, angular momentum along the axis of symmetry, and Carter's constant of the environment particles, normalized by their rest mass $\mu$. 

The number of particles per phase-space volume in the companion's local plane of simultaneity (i.e., $\bm{n}=\bm{u}$) is
\begin{equation}\label{eq:num_density}
	\frac{\dd N}{\dd V \dd^3v^i}\overset{*}{=} \gamma_v ^5 \, f(\mathcal{E},\mathcal{C},\mathcal{L}_z) \,,
\end{equation}
where, for later convenience, we changed coordinates in the momentum space to $v^a$, defined by $p^a\overset{*}{=}\mu \gamma_v v^a$, with~$\gamma_v~\equiv~(1-v^2)^{-\nicefrac{1}{2}}$ and~$v\equiv (v^a v_a)^{\nicefrac{1}{2}}$. The function $f(\mathcal{E},\mathcal{C},\mathcal{L}_z)$ depends on the type of environment and its evolution history, and can, in certain cases, be derived from first principles. At each position $x$, it is straightforward to relate $\mathcal{E}$, $\mathcal{C}$, and $\mathcal{L}_z$ to $v^i$, fixing the spatial frame vectors as $\bm{e}_1\propto (\bm{\partial}_r+u_r \bm{u})$, $\bm{e}_2\propto (\bm{\partial}_\theta+u_\theta \bm{u})$, and $\bm{e}_3\propto (\bm{\partial}_\varphi+u_\varphi \bm{u})$, where $(r, \theta, \varphi)$ are Boyer-Lindquist coordinates. The explicit expressions for quasi-circular EMRIs in Schwarzschild are given in the \textit{End Matter}. \\

{\em Interaction with environment.---}
%
In Fermi normal coordinates $(\tau, x^a)$ adapted to $\xi(I)$, with $I \equiv (\tau_1,\tau_1+\delta \tau)$, the tidal field from the primary BH vanishes at leading order in a neighborhood around $\xi(I)$~\cite{Manasse:1963zz}. Even so, when we ''zoom in'' close enough to the secondary object, its gravitational degrees of freedom become noticeable---this is often called the \emph{body zone}~\cite{Pound:2021qin_selfforce}; we expect the linear size of this region to be roughly of the order of the Hill radius, i.e., $x\lesssim x_{\rm H}\equiv R \,(q/3)^{\nicefrac{1}{3}}$, for a circular orbit of radius $R$ and a mass-ratio $q\equiv m/M\ll 1$. The extreme mass-ratio condition is crucial to establishing the separation of scales of the problem, which we will now explore.

For EMRIs immersed in incoherent collisionless media, we expect most of the energy-momentum exchange to happen locally inside the companion's body zone. Here, we will consider the companion to be a non-spinning BH interacting with the environment through gravity, but spin effects (e.g.,~\cite{Dyson:2024qrq, Wang:2024cej}) or extra couplings to matter may be added to our framework. In the body zone, the metric is well approximated by the Schwarzschild background (here in isotropic coordinates approaching asymptotically the Fermi normal coordinates)\footnote{We assume that the four-acceleration is small enough that its effects are negligible in the body zone.}
\begin{equation}
	\dd s^2\approx \frac{\left( 2\bar{r} - m \right)^2}{\left( 2\bar{r} + m \right)^2} \dd \tau^2 
	- \left( 1 + \frac{m}{2 \bar{r}} \right)^4\! \delta_{a b} \,\dd x^a \dd x^b
\end{equation}
with $\bar{r}\equiv (\delta_{a b} x^a x^b)^{\nicefrac{1}{2}}$. 

The four-acceleration arising from the gravitational interaction with the environment is (by definition)
\begin{equation}
	\bm{a}\equiv \nabla_{\bm{u}} \bm{u} = \frac{1}{m}\left[\nabla_{\bm{u}} (m \bm{u})-\frac{\dd m}{\dd \tau}\bm{u}\right]\,.
\end{equation}
The last expression can be easily evaluated in Fermi normal coordinates, using known results for particles scattering off a Schwarzschild BH. The linear momentum exchange is (in Fermi normal coordinates)
\begin{equation}\label{eq:accelera}
	\bm{a}\overset{*}{=}\frac{1}{m}\int \frac{\dd N}{\dd V \dd^3v^i} v \left[\sigma_{\rm acc}(v) + \sigma_{\rm scat}(v)\right]\bm{p}_\perp \dd^3 v^i\,,
\end{equation}
with $\bm{p}_\perp\overset{*}{=}\mu\gamma_v(0,v^a)$, $\dd N/(\dd V \dd^3 v^i)$ as in~\eqref{eq:num_density}, and $\sigma_{\rm acc}$ and $\sigma_{\rm scat}$, respectively, the accretion (acc) and gravitational scattering (scat) cross sections. In the last expression, we used the isotropy of Schwarzschild.\footnote{The companion's spin (or asphericity) will increase the complexity of the problem, but does not pose an obstacle to our approach.} 
The integral in \linebreak $\dd v^3\equiv \dd v^1 \dd v^2 \dd v^3$ will generally have to be performed numerically.

The acceleration computed in Fermi normal coordinates adapted to the companion has then to be related to the rate of change of the conserved quantities describing its motion. Explicit expressions for quasi-circular EMRIs in Schwarzschild are given in the \textit{End Matter}. \\

{\em Cold dark matter overdensity.---}
For concreteness, we focus on EMRIs in the presence of cold and collisionless DM. The density and velocity distribution of DM particles around BHs has been extensively discussed in the literature (see~\cite{Bertone:2024rxe} for a recent review). If BHs grow adiabatically from an infinitesimal seed at the center of a DM halo with a power law density profile, DM is expected to develop a dense DM ``spike'' characterized by a density profile which is again a power law, but with a steeper slope~\cite{Quinlan:1994ed, Gondolo:1999ef}. In practice, the finite mass and size of BH seeds imply that within a radius equal to the size of the star that eventually collapses to the BH seed, the DM distribution is predicted to form an overdensity shallower than a DM spike, referred to as DM ``mound''~\cite{Bertone:2024wbn}. A fully relativistic treatment of the evolution of the DM distribution function also leads to a substantial deviation of the DM density profile from the case of a Newtonian DM spike, both in the case of a Schwarzschild~\cite{Sadeghian:2013laa} and a Kerr BH~\cite{Ferrer:2017xwm}. 

We consider a relativistic DM spike around a massive Schwarzschild BH, and derive the distribution $f(\mathcal{E},\mathcal{C},\mathcal{L}_z)$ from adiabatic invariants as in~\cite{Sadeghian:2013laa}.\footnote{As noted in footnote~3 of~\cite{Ferrer:2017xwm}, there is a $(1-2M/r)^{-1}$ factor missing in the Schwarzschild's radial invariant in Eq.~(3.19) of~\cite{Sadeghian:2013laa}.} We consider a Milky-Way-like galaxy, assuming an isotropic host DM halo described by a Hernquist density profile~\cite{Hernquist:1990be} with mass $M_{\rm h}=10^{12} M_\odot$ and scale length $a=20\,{\rm kpc}$, and a supermassive BH with mass $M=10^6 \,M_\odot$. The resulting DM spike profile has a peak rest mass density of $2.4 \times 10^{19}\,\mathrm{GeV/cm^3}$ at $r\approx 7.7 M$, as measured by stationary observers with four-velocity $\bm{n}=(-g_{tt})^{-\nicefrac{1}{2}}\bm{\partial}_t$; this scales approximately with $\propto M^{-1.7}$~(e.g.,~\cite{Speeney:2022ryg}). \\

{\em Particle dark matter.---}
%
The process of collisionless DM scattering off a Schwarzschild BH companion is determined completely by the spacetime geodesic structure.
As shown in App.~B of~\cite{2305.10492}, the cross sections in~\eqref{eq:accelera} are $\sigma_{\rm acc}=\pi b_{\rm cr}^2(v)$ and $ \sigma_{\rm scat}=2\pi \int_{b>b_{\rm cr}} \dd b\,b[m \cos[\phi_\infty(b,v)]/b_{\rm gr}(v)]^2 $, where $b_{\rm gr} \equiv (m/v^2) (1+v^2)$ and the critical (capture) impact parameter, $b_{\rm cr}$, is expressed analytically in~\cite{2305.10492}.
The asymptotic true anomaly~$\phi_\infty$ can be expressed in terms of elliptic integrals of the first kind~\cite{chandrasekhar1998mathematical, 2305.10492}. 
\begin{figure}[t]
\centering
\hspace*{-0.05\columnwidth}
    \includegraphics[width=\columnwidth]{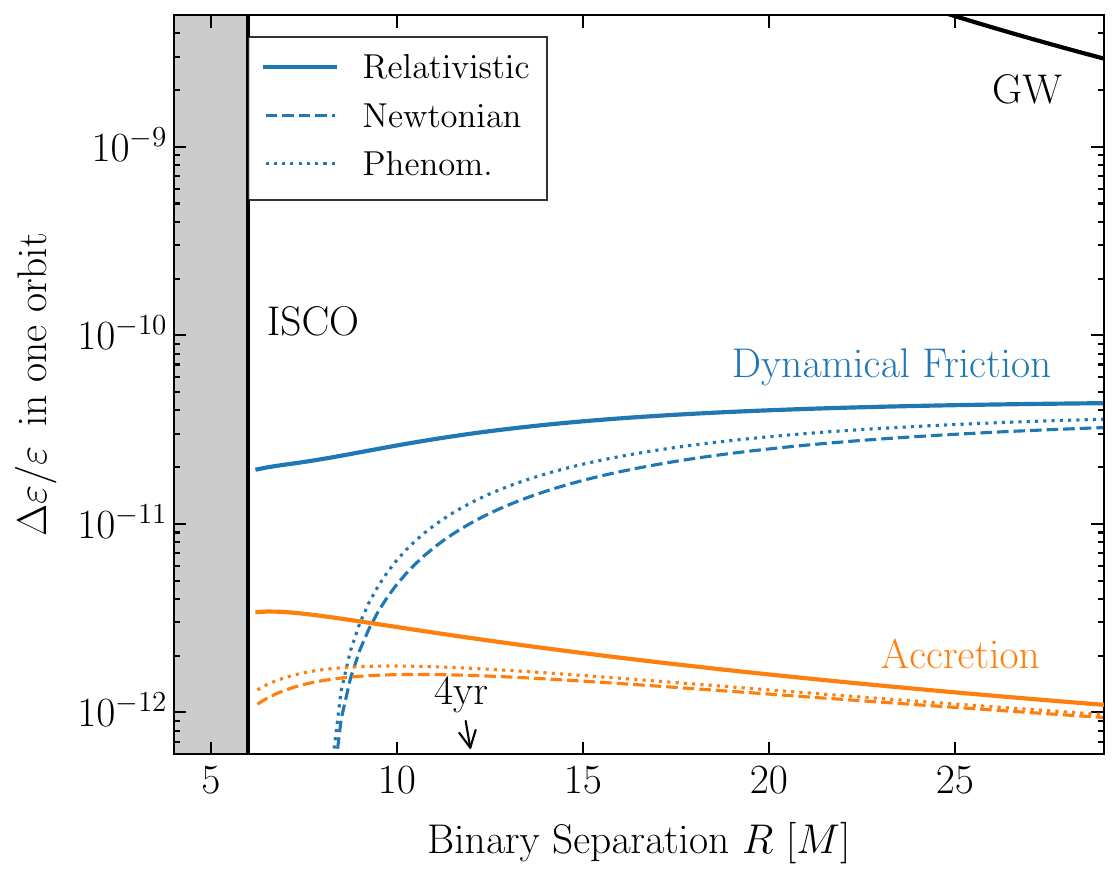}
    \caption{\textbf{Energy loss per orbital cycle as a function of binary separation.} Solid lines correspond to the fully relativistic model, the dashed to the Newtonian, and the dotted to including heuristic relativistic corrections as in~\cite{Speeney:2022ryg}, for an EMRI with mass-ratio $q=10^{-5}$ in a DM spike of a Milky-Way-like galaxy. An EMRI at $R = 12 M$ will merge after $\sim4$ years.\label{fig:losses}}
\end{figure}

The scattering cross section can be made more explicit by expressing $\sigma_{\rm scat}\equiv 4\pi \Lambda \, b_{\rm gr}^2$, introducing the Coulomb logarithm (using an infrared cutoff $b_{\rm max}\sim x_{\rm H}$)
\begin{equation*}
	\Lambda(v) = \ln\sqrt{\tfrac{b_{\rm max}^2+b_{\rm gr}^2(v)}{b_{\rm cr}^2(v)+b_{\rm gr}^2(v)}}+\chi(v)\, ,
\end{equation*}
which includes the relativistic contribution
\begin{equation*}
	\chi \equiv \ln\sqrt{\tfrac{b_{\rm cr}^2(v)+b_{\rm gr}^2(v)}{b_{\rm min}^2+b_{\rm gr}^2(v)}}+ b_{\rm gr}^{-2}(v) \, \int_{b_{\rm cr}(v)}^{b_{\rm min}}\cos^2 [\phi_\infty(b,v)] b\, \dd b\,,
\end{equation*}
where~$b_{\rm min}$ is chosen freely, provided $b_{\rm cr}\ll b_{\rm min}\leq b_{\rm max}$. The function~$\chi$ is monotonically increasing and~$O(1)$ for~$v\gtrsim10^{-1}$. It is accurately approximated by a simple analytic expression given in the Supplemental Material~\cite{SupplementalMaterial}.

Figure~\ref{fig:losses} shows the fractional energy loss during an orbit, with $\Delta\varepsilon = \dot{\varepsilon} \,(2\pi/\Omega)$, due to momentum accretion, dynamical friction (elastic scattering), and gravitational wave radiation as a function of the binary separation $R$, computed through~\eqref{eq:rates} by numerically integrating over the velocity space in~\eqref{eq:accelera} and using $b_{\rm max}=x_{\rm H}$. We compare the results against a Newtonian model of the environmental effects and one including phenomenological relativistic corrections as in~\cite{Speeney:2022ryg}. The curves show that both models largely underestimate dynamical friction.

In both models, we keep the relativistic expressions \eqref{eqs:integrals_DF} in the distribution function, to preserve the loss cone of the DM distribution; a Newtonian modeling of the loss cone would result in a depleted region $r<8 M$ \cite{Gondolo:1999ef} (as opposed to the actual $r<4 M$ \cite{Sadeghian:2013laa}). We also keep $b_{\rm cr}(v)$ to compute the effect of accretion, despite that gravitational capture is a relativistic effect. We neglect the relativistic factors $\gamma_v$ in the incoming momentum $\bm{p}_\perp$, $(1+v^2)$ in the scattering impact parameter $b_{\rm gr}$, $\chi$ in the Coulomb logarithm $\Lambda$, the strong-field terms in the specific torque in \eqref{eq:rates}, and use $\dd N/(\dd V \dd^3v^i)$ as measured by the stationary observers [$\bm{n}=(-g_{tt})^{-\nicefrac{1}{2}}\bm{\partial}_t$] in the integration \eqref{eq:accelera}. The latter (multiplied by the rest mass $\mu$) coincides with the mass density obtained in~\cite{Sadeghian:2013laa}, which has been used in previous models of environmental effects. 

\begin{figure}[t!]
\centering
\hspace*{-0.05\columnwidth}
    \includegraphics[width=\columnwidth]{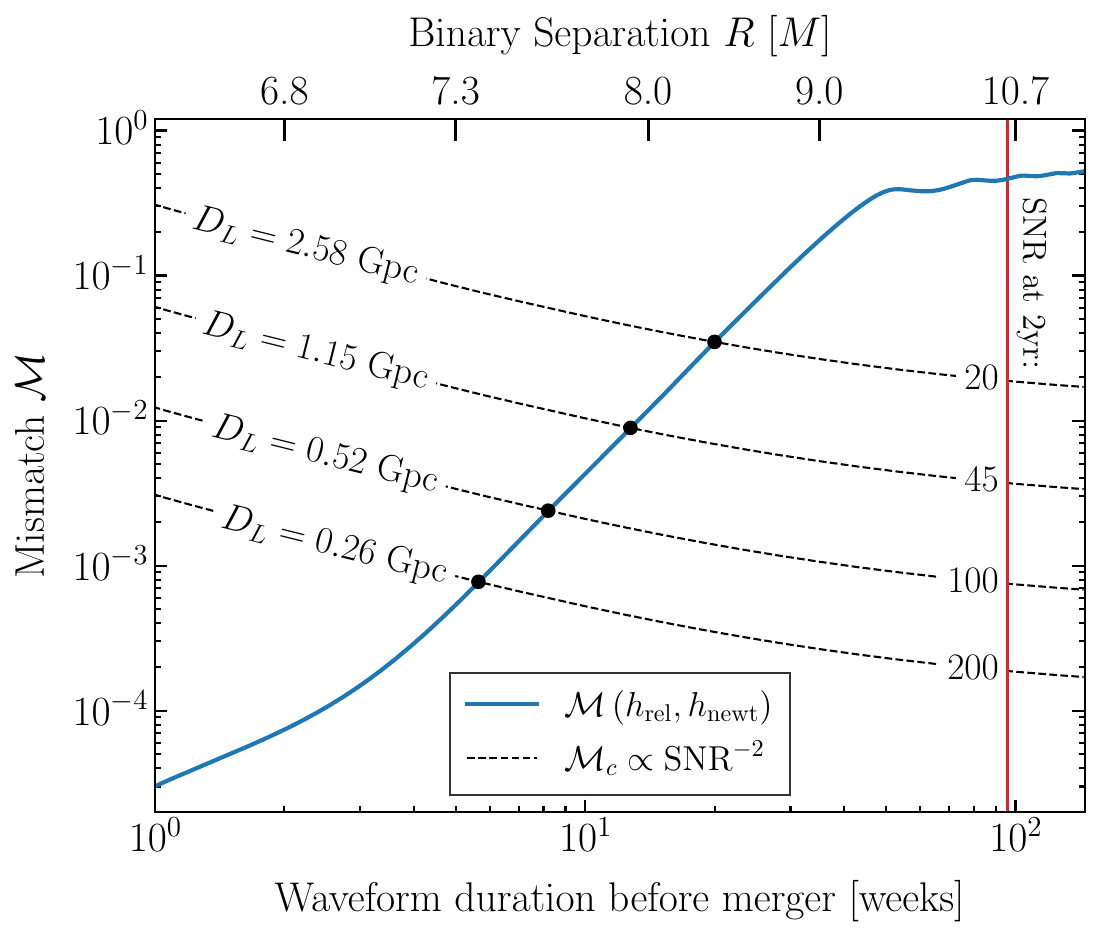}
    \caption{\textbf{Mismatch between relativistic and Newtonian environmental models as a function of observation time (in weeks before merger).} 
    Dashed lines mark the mismatch thresholds for various luminosity distances $D_L$, above which the Newtonian environmental waveform diverges significantly from the actual relativistic one, introducing systematics into LISA measurements. For an EMRI with mass-ratio $q=10^{-5}$ embedded in a dark-matter spike of a Milky-Way-like galaxy, the mismatch exceeds the critical threshold after only a few weeks of observation.
    \label{fig:mismatch_duration}}
\end{figure}

For comparison, we also consider a phenomenological model of the relativistic corrections motivated by~\cite{Speeney:2022ryg} (labeled ``Phenom.''). This includes a multiplicative factor $\gamma_u^2\equiv1/(1-u^{2})$, with $u\equiv \Omega R$, to the dynamical friction and accretion torques, as an attempt to account for the volume contraction and the relativistic correction to the particles momentum from the companion's perspective, and also a multiplicative factor~$(1+u^{2})^{2}$ to the dynamical friction torque, from relativistic corrections to $b_{\rm gr}^2$ (cf. previous paragraph). 
The deviation from the fully relativistic model (shown in Fig.~\ref{fig:mismatch_duration}) arises from neglecting strong-field effects of both the supermassive and companion BHs, and from considering the (heuristic) relativistic corrections out of the integral over the velocity space~\eqref{eq:accelera} (see Supplemental Material~\cite{SupplementalMaterial}).\\

{\em Impact of relativistic effects.---}
%
We quantify the difference between two waveform-modeling frameworks using the ``faithfulness'' integral $\mathcal{F} \equiv \left<a| b\right>/\left(\rho_a \rho_b\right)$~\cite{Damour:1997ub} and its complement, the ``mismatch'' $\mathcal{M} = 1 -\mathcal{F}$, where the noise-weighted inner product between two real waveforms $a(t)$ and $b(t)$ is given in frequency domain by the usual form (e.g.,~\cite[Eq.~(2.3)]{Cutler:1994ys}),
with the one-sided LISA power-spectral density taken to be the numerical Michelson-like PSD from~\cite{Robson_2019}, and $\rho_a~\!\!\!\equiv~\!\!\!\sqrt{\left<a| a\right>}$ is the optimal signal-to-noise-ratio (SNR). We evaluate the noise-weighted inner product from $f_\mathrm{min}~\!\!=~\!\!10^{-5}$~Hz up to the Nyquist frequency $f_\mathrm{max} = 0.05$~Hz, and maximize the result over phase and time offsets as in~\cite{Coogan:2021uqv}. We obtain the EMRI waveforms in each case by incorporating Eqs.~\ref{eq:rates} into the \texttt{FastEMRIWaveforms\_v1.5.4} (FEW) framework~\cite{Katz:2021yft,Chua_2021,Speri_2024,chapmanbird2025fastframedraggingefficientwaveforms}, initializing binaries 4\,yr before merger.

Figure~\ref{fig:mismatch_duration} characterizes the importance of relativistic effects in the modeling of environmental effects in terms of what future mHz interferometers like LISA can discern. 
The ability of the detector to distinguish two waveforms grows with the observation time as the SNR accumulates.
In particular, one can estimate the critical mismatch above which two waveforms can be distinguished in the detector by~$\mathcal{M}_c = (n/2)\,\mathrm{SNR}^{-2}$~\cite{Chatziioannou:2017tdw}, where~$n$ is the number of intrinsic parameters (here, equal to~$15$). The critical mismatch is shown in Fig.~\ref{fig:mismatch_duration} as a function of the observation time for different luminosity distances (or optimal SNR at $2$ yr of observation). For periods of observation with $\mathcal{M}>\mathcal{M}_c$, the relativistic effects must be included in the modeling of environmental effects to avoid systematic biases or risk a substantial loss in SNR. As a reference, for our benchmark system with $q=10^{-5}$, this observation period falls roughly between 1 and 4.5 months for luminosity distances $D_L\in (0.3,2.6)\,{\rm Gpc}$. While the precise values of the mismatch between the relativistic and Newtonian models may vary with the EMRI parameters, we do not expect our qualitative conclusions to be affected.

Newtonian models largely underestimate the impact of environmental effects on EMRI waveforms. For our benchmark system with mass-ratio $q=10^{-5}$, we verified that an observation of more than a year before merger is required for LISA to discern those models from vacuum, as opposed to the few weeks for the fully relativistic model. Our results also show that relativistic corrections to environmental effects are comparatively weaker for systems with less extreme mass ratios, which can be understood from the fact that such systems chirp faster through the strong-field region.  \\

{\em Discussion.---}
%
In this Letter, we introduce a fully relativistic framework to include environmental effects of collisionless environments in the evolution of EMRIs. Applying it to the case of DM spikes around massive BHs, we showed that a fully relativistic treatment is necessary to accurately describe the environmental dephasing on such signals.
In this work, we neglected the effect of the EMRI on the DM overdensity. Using the Newtonian framework \texttt{HaloFeedback}~\cite{HaloFeedback, Kavanagh:2020cfn}, we confirmed that an EMRI with mass-ratio $q=10^{-5}$ would lead to tiny feedback on the DM overdensity: the evolution of the distribution function would lead to less than one cycle of dephasing in 4 yr (corresponding to $\sim 0.5\%$ of the total dephasing). 
We note, though, that, for LISA data analysis, post-adiabatic EMRI waveforms with errors below one radian dephasing after 4 yr are required.

\begin{figure}[t!]
\centering
\hspace*{-0.05\columnwidth}
    \includegraphics[width=\columnwidth]{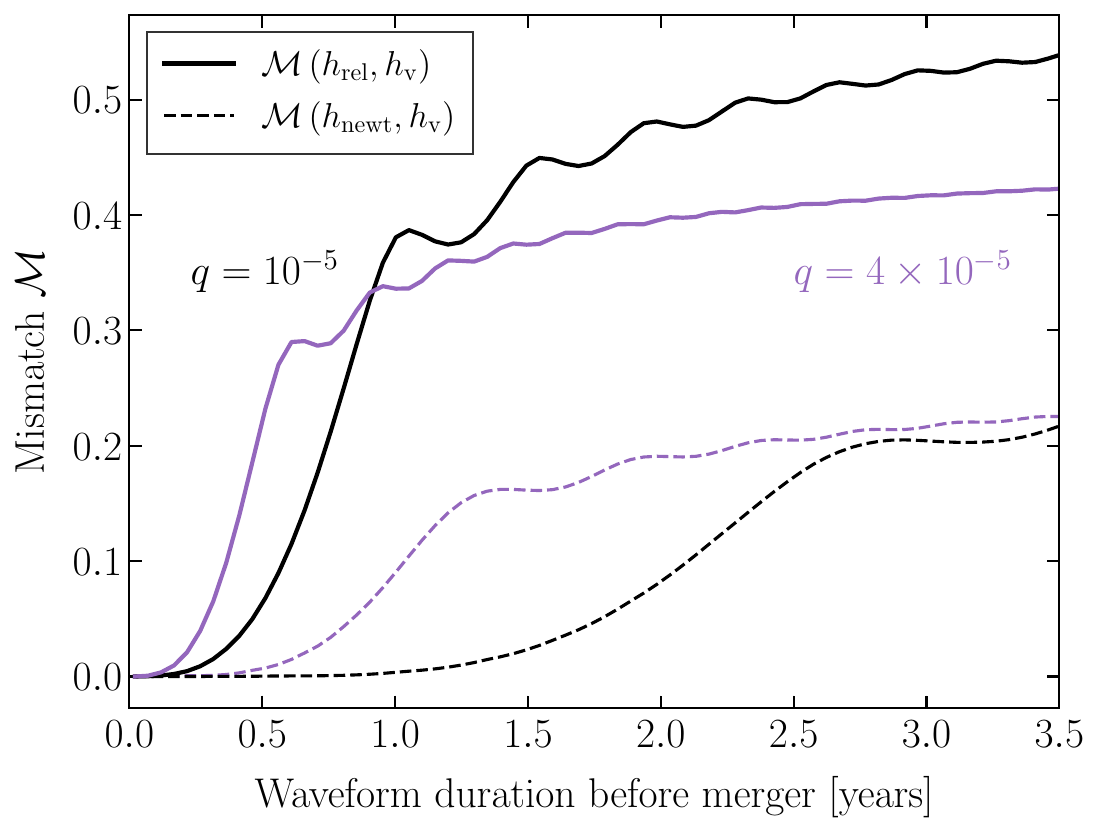}
    \caption{\textbf{Mismatch with respect to vacuum waveforms as a function of signal duration (in years before merger).} The black and purple lines are for the benchmark $(q=10^{-5})$, or a slightly heavier companion.\label{fig:mismatch_q}}
\end{figure}

A different route to relativistic environmental effects on the generation and propagation of GWs was taken in~\cite{Cardoso:2021wlq} (with subsequent works~\cite{Cardoso:2022whc, Spieksma:2024voy, Speeney:2024mas, Pezzella:2024tkf, Gliorio:2025cbh}), which includes the self-gravity of surrounding matter. This was based on a result by~\citet{Einstein:1939ms}, that a system of many gravitating point masses moving along circular geodesics about a common center can be mapped into an effective anisotropic fluid with only tangential pressure. Because real DM halos should contain particles distributed over a wide range of eccentricities due to processes like violent relaxation~\cite{Lynden-Bell:1967, Spergel:1992}, the physical interpretation of results derived for ``Einstein clusters'' should be done with care. Very rich phenomenology is observed to emerge from the coupled (linear) perturbations in the gravitational-matter sectors in such approaches. Typical halos, as our benchmark Milky-Way-like halo, have dimensionless compactness $\lesssim 10^{-6}$, leading to very small couplings between gravity and matter perturbations; it is conceivable, though, that such small effects might accumulate over the large halo scales into observable imprints in GW waveforms. In low-compactness collisionless halos, effects on GW propagation described within geometric optics approximation via a gravitational redshift~\cite{Laguna:2009re} lead roughly to one dephasing cycle after $\sim10^6$ cycles; this is $\sim 10^3$ times smaller than the effects of dynamical friction studied here (as found in~\cite{Mitra:2025tag}).

In this Letter, we imposed a hard cutoff on the impact parameter, (conservatively) neglecting the interactions with DM particles outside the secondary's Hill radius, where the validity of our assumptions breaks down. We expect the total torque resulting from additional gravitational three-body interactions to increase the impact of environmental effects on EMRI evolution. Such ``stirring'' effect could, in principle, be obtained perturbatively by computing the energy and angular momentum exchange with the DM particles mediated by (vacuum) linear metric perturbations, similarly to the approach of~\cite{2501.09806} to ultralight scalar fields.

Direct extensions of this work include the application of our framework to eccentric and inclined EMRIs onto a primary Kerr BH. We note that cold DM spikes can be significantly denser around spinning BHs~\cite{Ferrer:2017xwm}. Moreover, we expect relativistic effects in the interaction of EMRIs with environments to be even more important for prograde orbits around highly spinning BHs, as their ISCO (and marginally bound orbits) approaches the event horizon radius. Here, we neglected the spin of the secondary, but the results of~\cite{Dyson:2024qrq} could be directly incorporated into our framework. Surrounding environments are responsible for a characteristic coevolution of the mass and spin of the companion (explored in~\cite{Karydas:2025bkj}); while unmeasurable for quasicircular EMRIs~\cite{Piovano:2021iwv}, the spin of the companion might be probed with eccentricity~\cite{Cui:2025bgu}.

\vspace{0.5 cm}
{\em Acknowledgments.---}
%
We thank Francisco Duque and Lorenzo Speri for valuable discussions and guidance on implementing our results in FEW. We are also grateful to Miguel R. Correia for insightful conversations in the early stages of this work. We also thank Emanuele Berti, Vitor Cardoso, Francisco Duque, Andrea Maselli, and Nick Speeney for their helpful comments and stimulating discussions on our draft. We gratefully acknowledge the support of the Dutch Research Council (NWO) through an Open Competition Domain Science-M grant, project number OCENW.M.21.375. 

%

\vspace{2pt}
\section{End Matter}
%
{\em Quasi-circular EMRIs in Schwarzschild.---}
%
For quasi-circular EMRIs in Schwarzschild ($r=R$, $\theta=\pi/2$, $u_\theta=u_r=0$), the conserved quantities of the DM particles can be expressed in terms of their velocity in the companion’s proper frame as
\begin{equation}
\begin{gathered}\label{eqs:integrals_DF}
	\frac{\mathcal{E}}{\varepsilon}\overset{*}{=} \gamma_v \frac{1+v_3(l_z/R)}{\sqrt{1+(l_z/R)^2}}\,,\qquad \mathcal{C}^{\nicefrac{1}{2}}\overset{*}{=}\gamma_v R \,|v_2|\,,\\ \frac{\mathcal{L}_z}{l_z} \overset{*}{=}  \gamma_v \left[1+v_3 \sqrt{1+(R/l_z)^2} \right] \,,
\end{gathered}
\end{equation}
where $\varepsilon=\frac{R-2M}{\sqrt{R(R-3M)}}$ and $|l_z|=\frac{R \sqrt{M}}{\sqrt{R-3M}}$ are the companion’s specific energy and angular momentum.

The rate of change of the conserved quantities of a (equatorial) quasi-circular EMRI in Schwarzschild will be [from Eqs.~\eqref{eq:rates0}]
\begin{equation}
\begin{gathered}\label{eq:rates}
	\frac{\dot{\varepsilon}_e}{\varepsilon}\overset{*}{=}\frac{ a_3 \left(l_z/R\right) }{\sqrt{1-\frac{3M}{R}}\sqrt{1+(l_z/R)^2}} \,, \quad  \dot{C}_e^{\nicefrac{1}{2}}\overset{*}{=} \frac{R \, |a_2|}{\sqrt{1-\frac{3M}{R}}}\,, \\ 
	\frac{\dot{l}_{z,e}}{|l_{z}|}\overset{*}{=}\frac{a_3}{\sqrt{1-\frac{3M}{R}}}\sqrt{(R/l_z)^2+1}\,,
\end{gathered}
\end{equation}
where the acceleration components are in the companion's proper frame.
The orbit is maintained equatorial if $a_2$ (or its orbit average) vanishes. In particular, this is the case if the environment's distribution function is symmetric with respect to the equatorial plane.

{\em Parallel transport of $\bm{e}_a$.---}
%
The careful reader may have noticed that the frame $\bm{e}_a$ chosen in the previous section is not parallel transported along $\xi$. This means that, in Fermi normal coordinates, the expressions~\eqref{eqs:integrals_DF}---which we use to compute the integral over the distribution function in~\eqref{eq:accelera}---are strictly valid only at a given instant $\tau_1$. Nevertheless, they remain a good approximation for an interval $\delta \tau\ll R^2/|l_z|$, the timescale for the parallel transported frame to change significantly from the one we use. As the gravitational scattering happens over a very much shorter timescale $\delta \tau_{\rm int} \lesssim (x_{\rm H}/R)\sqrt{1+(l_z/R)^2} (R^2/|l_z|) \ll (R^2/|l_z|)$,\footnote{We used $\delta \tau_{\rm int}\lesssim x_{\rm H}/|\braket{v^3}|$, assuming $\braket{\mathcal{L}_z}\approx 0$; but the argument holds as long as $|\braket{v^3}|\sim |l_z|/R$.} we can to a good approximation use~\eqref{eqs:integrals_DF} to evaluate \eqref{eq:accelera}. \\

{\em Heuristic corrections.---}
%
In Fig.~\ref{fig:mismatch_duration} we compared the energy losses from our self-consistent relativistic treatment of environmental effects in DM spikes with those from a heuristic relativistic prescription that still accounts for the DM velocity dispersion (as $\sigma_{\rm acc}$ and $\sigma_{\rm scat}$ are integrated over velocity space). The reasons for the large differences observed were discussed in the text.
Here, we report that in~\cite{Speeney:2022ryg} the Coulomb logarithm was fixed to the (arbitrary) constant value $\Lambda=3$, and the velocity distribution was not taken into account when computing the acceleration; coincidentally, this results in a torque closer to our fully relativistic model.

\section{Supplemental Material}

{\em Relativistic Coulomb logarithm.---}
%
In computing the momentum transferred by particles scattering off a Schwarzschild spacetime, the contribution from orbits with asymptotic velocity $v$ and impact parameter $b\gtrsim b_{\rm cr}(v)$ (i.e., only slightly larger than the one of capture) must be computed numerically~\cite{Traykova:2023qyv}. This contribution can be encapsulated in the factor $\chi(v)$ defined in the main text. Choosing~$b_{\rm min}=40 b_{\rm cr}(v)$, the function~$\chi$ is accurately described (with an error $\leq 1.5 \%$) by the fit
\begin{equation*}
	\chi \approx \ln\sqrt{\tfrac{b_{\rm cr}^2(v)+b_{\rm gr}^2(v)}{b_{\rm min}^2+b_{\rm gr}^2(v)}}+ 12786 v^2\left(\frac{1-0.096 v^3}{1+1844 v^{1.62}}\right)\,.
\end{equation*}
\\

\begin{figure}[t!]
\centering
\hspace*{-0.05\columnwidth}
    \includegraphics[width=\columnwidth]{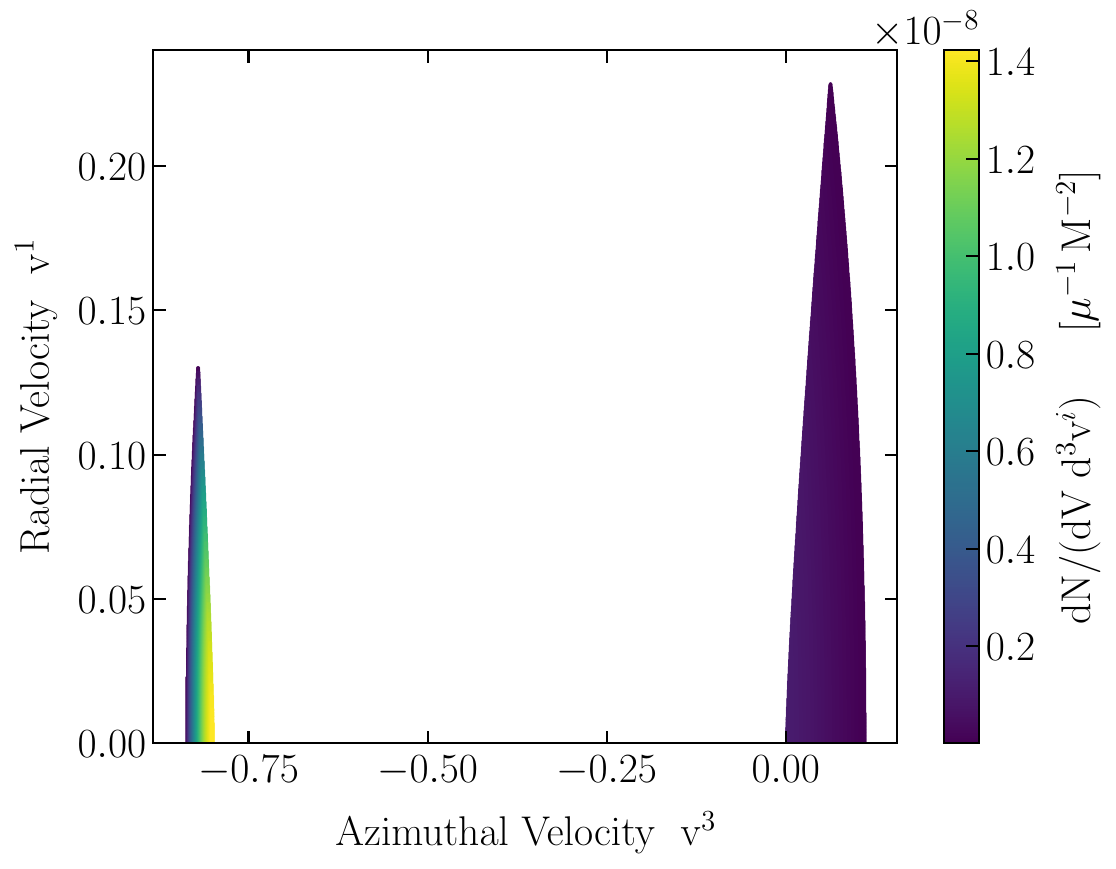}
    \caption{\textbf{Distribution function in the companion frame at the ISCO (\boldmath$R=6 M$) for the region \boldmath$v^1\geq 0$ of a slice \boldmath$v^2=0$.} For a (quasi)circular EMRI, the distribution function is symmetric under $v^1\to-v^1$. The white region shows the loss cone of our fiducial DM spike; only the velocities in the colored regions correspond to stable bound orbits around the supermassive BH. \label{fig:phase_space}}
\end{figure}

{\em DM distribution function in velocity space.---}
%
To better understand the distribution of velocities in the companion's frame, in Fig.~\ref{fig:phase_space} we show a density plot of the distribution function $\dd N/(\dd V\dd^3 v^i)$ for a companion in a circular orbit\footnote{Or at pericenter or apocenter.} at the ISCO radius. We show the region $v^1\geq 0$ of a slice $v^2=0$, as the distribution function is symmetric under $v^1\to-v^1$ (for equatorial orbits, it is also symmetric under $v_2\to - v^2$).

The white region represents the loss cone of the DM spike, i.e., only particles with velocities within the colored regions are in stable bound orbits around the supermassive BH. All the others are either unbound or captured by the supermassive BH. We notice two disconnected regions associated with DM particles in prograde and retrograde motion relative to the companion orbit. For a companion in circular orbit at large radii, the prograde region is roughly centered at $v^3=0$, while for orbital radii in the strong-field, most prograde orbits have $v^3<0$, contributing a net positive torque to the EMRI. Simplified models that neglect the velocity anisotropy of the distribution cannot capture the radius-dependent competition between positive and negative torques, which can affect the final torque. In fact, when using the distribution function $\dd N/(\dd V \dd^3v^i)$ as measured by the stationary observers [$\bm{n}=(-g_{tt})^{-\nicefrac{1}{2}}\bm{\partial}_t$] in the integration in Eq.~(7), we observe a change of sign in the friction force happening at $r \approx 8M$ (close to the peak density). Thus, the `friction' effectively becomes an `impulse' at smaller radii than $8M$. We have also checked that the same happens for a Newtonian computation with Gondolo-Silk spike~\cite{Gondolo:1999ef}; in that case, the transition happens at a larger radius $r \approx 15M$ (again near the peak density).

As explained in the text, both in the Newtonian and phenomenological (with heuristic relativistic effects) models, we used the distribution function perceived by stationary observers $\bm{n}=(-g_{tt})^{-\nicefrac{1}{2}}\bm{\partial}_t$ to evaluate the integral in $\dd^3 v^i$. Their different local plane of simultaneity changes the distribution function relatively to the companion's (shown in Fig.~\ref{fig:phase_space} for $R=6M$) by a multiplicative factor $(-g_{tt})^{-\nicefrac{1}{2}}\mathcal{E}/\gamma_v$, while not changing the loss cone. This implies that these observers perceive a considerably smaller density of particles in retrograde orbits and slightly larger in prograde orbits.\\

\begin{figure}[b]
	\centering
	\hspace*{-0.05\columnwidth}
	\includegraphics[width=\columnwidth]{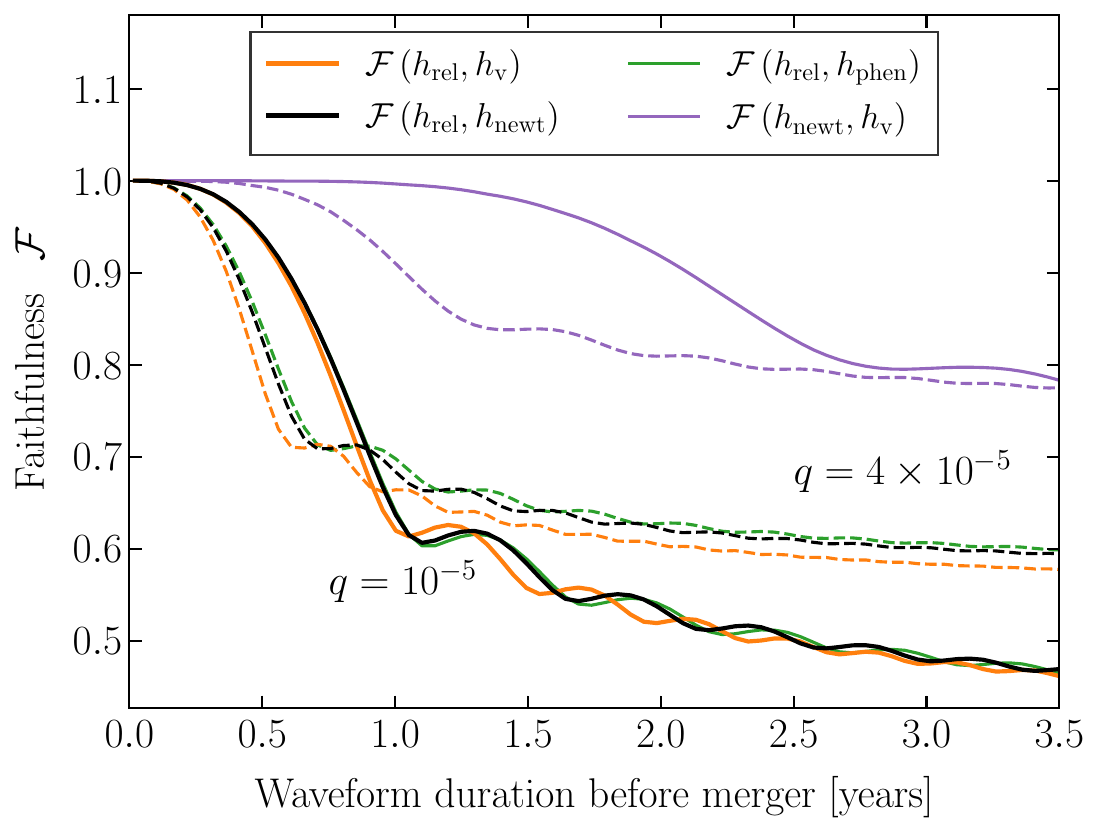}
	\caption{\textbf{Faithfulness between different models as a function of signal duration (in years before merger).} The solid lines are for the benchmark $q=10^{-5}$ system, while the dashed are for a slightly heavier companion.\label{fig:faithfulness}}
\end{figure}

{\em Comparison between different models.---}
%
In Fig.~\ref{fig:faithfulness}, we compare the faithfulness between multiple models as a function of observation period. 
The oscillations are due to the incidental interference for long waveforms and do not carry any physical meaning (see, e.g.,~\cite{2410.17310}).
We show that the Newtonian and phenomenological models (with heuristic relativistic effects) can largely underestimate the impact of environmental effects on GW waveforms. It is also shown that relativistic corrections to interaction with the environment are comparatively weaker for systems with less extreme mass ratios, as such systems chirp faster through the strong-field region.

\end{document}